Improving clinical trial interpretation with ACCEPT analyses

Michelle N. Clements[1], Ian R. White[1], Andrew J. Copas[1], Victoria Cornelius[2], Suzie Cro[2], David T Dunn[1], Matteo Quartagno[1], Rebecca M. Turner[1], Conor D. Tweed[1], A. Sarah Walker[1]

[1]MRC Clinical Trials Unit at UCL, London, UK

[2]Imperial Clinical Trials Unit, London, UK

Effective decision making from randomised controlled clinical trials relies on robust interpretation of the numerical results. However, the language we use to describe clinical trials can cause confusion both in trial design and in comparing results across trials. ACceptability Curve Estimation using Probability Above Threshold (ACCEPT) aids comparison between trials (even where of different designs) by harmonising reporting of results, acknowledging different interpretations of the results may be valid in different situations, and moving the focus from comparison to a pre-specified value to interpretation of the trial data. ACCEPT can be applied to historical trials or incorporated into statistical analysis plans for future analyses. An online tool enables ACCEPT on up to three trials simultaneously.

The classic superiority trial aims to generate robust evidence that a new treatment is better than placebo. Active controls are used to assess superiority of a new treatment when the use of placebo is unethical, such as when an effective treatment is available. Non-inferiority trials, aiming to show that the new treatment is not appreciably worse than the control, are often used to evaluate new drugs and interventions with similar expected efficacy to standard therapy but secondary advantages such as less toxicity, ease of implementation, benefits in particular subgroups only, or lower cost. Similarly, equivalence trials aim to show a new treatment is unlikely to differ appreciably from control in either direction and super-superiority trials aim to show evidence of a new treatment being better than control by at least a specified value.

The specification of trial type (e.g. as superiority or non-inferiority) is important to enable assessment of whether a trial has met its aims. However, the different terms are in themselves confusing. Additionally, seemingly paradoxical situations can arise when comparing across trials, such as trial type differing depending on which treatment is assigned as "control", and trials with similar numerical results reaching different conclusions.

All clinical trial types can be linked by the pre-specified 'unacceptable value' that the 95% confidence interval limits of the estimate of the difference between treatments are compared to. Specification of trial type is equivalent to pre-specification of the unacceptable value: zero (or one for a relative effect measure) in superiority trials, the non-inferiority margin (less than zero) in non-inferiority trials and greater than zero in super-superiority trials.

Comparison to the pre-specified unacceptable value is an appropriate part of trial interpretation but leads to a binary conclusion of 'trial aim met' or 'trial aim not met'. Outside drug regulation, binary conclusions are widely viewed as problematic[1], as evidence should not be reduced to a single threshold but should be considered in context with other factors such as the point estimate and confidence interval[2]. Interpretation of non-inferiority trial results also suffers from added complexity around whether the pre-set non-inferiority margin was justified or is relevant for settings outside the trial. Importantly, stakeholders such as clinicians, patients or policy makers may have differing but equally valid unacceptable difference values depending on the relative importance placed on secondary factors such as cost or toxicity.

We advocate the wider use of ACCEPT as secondary analyses in clinical trials. We illustrate this using two HIV trials, EARNEST[3] and SECOND-LINE[4], which had similar quantitative results but drew different conclusions5. We demonstrate how alternative presentation of results could aid better comparison and integration of their findings. We present the trials together, but imagine ACCEPT being presented in each trial results paper separately as secondary analyses. For ease throughout we measure differences as treatment minus control for a favourable outcome, so that positive values indicate higher efficacy in the tested treatment.

EARNEST and SECOND-LINE investigated raltegravir as a second-line therapy for HIV in comparison to standard therapy of nucleoside reverse transcriptase inhibitors (NRTI). EARNEST, was carried out in low and middle-income countries. It was pre-specified as a superiority trial because raltegravir was more expensive than NRTI: so it was thought that clear benefit would have to be shown for implementation. The pre-specified unacceptable value was consequently zero.

SECOND-LINE, was carried out predominantly in high-income countries. It was pre-specified as a non-inferiority trial because raltegravir was considered to have a better toxicity profile than NRTI: implementation was therefore considered to be worthwhile with similar efficacy. The pre-specified unacceptable value was the non-inferiority margin of minus 12% on the risk difference scale.

The original analysis of EARNEST compared the lower limit of the 95% CI of the difference between treatments (-2.4%) to the unacceptable value of 0, drawing the conclusion of 'superiority not shown' and implementation was not recommended. Analysis of SECOND-LINE compared the lower limit of the 95% CI (-4.7%) to the unacceptable value of -12%, drawing the conclusion of 'non-inferiority' and implementation was recommended. The question then arises of how two trials with numerically similar results can reach opposing conclusions regarding implementation.

Differing, valid, opinions on the unacceptable differences values (0% in EARNEST and -12% in SECOND-LINE), driven in part by different emphasis on secondary benefits, led to the selection of different trial types and the resulting seemingly opposing recommendations. Interpretation through ACCEPT, including both graphs and tables, would have helped to clarify this paradox enabling more nuanced interpretation of the results.

ACCEPT uses the primary analysis from a trial to plot the probability of the true difference between treatments being above an 'acceptability threshold' for a range of possible threshold values (Figure 1). ACCEPT can be presented for all trial types and outcomes. ACCEPT has only been used sporadically in clinical trials 6-12 with no common naming. ACCEPT is similar to cost-effectiveness acceptability curves widely used in health economics, where weight of evidence, rather than binary conclusions, is a more widely accepted paradigm.

ACCEPT output is best presented in a graph with associated tables. A graph shows a continuous range of acceptability thresholds where greater uncertainty around point estimates (with larger associated confidence intervals) is reflected in a shallower slope. Additional tables present selected acceptability thresholds or the probability that the true value is between selected thresholds. To enable comparison of ACCEPT between trials, tables should include acceptability values for the unacceptable difference (specified in the trial design), zero, a reasonable range of potential alternative unacceptable values, and acceptability thresholds for the 2.5th, 50th, and 97.5th percentile acceptability values.

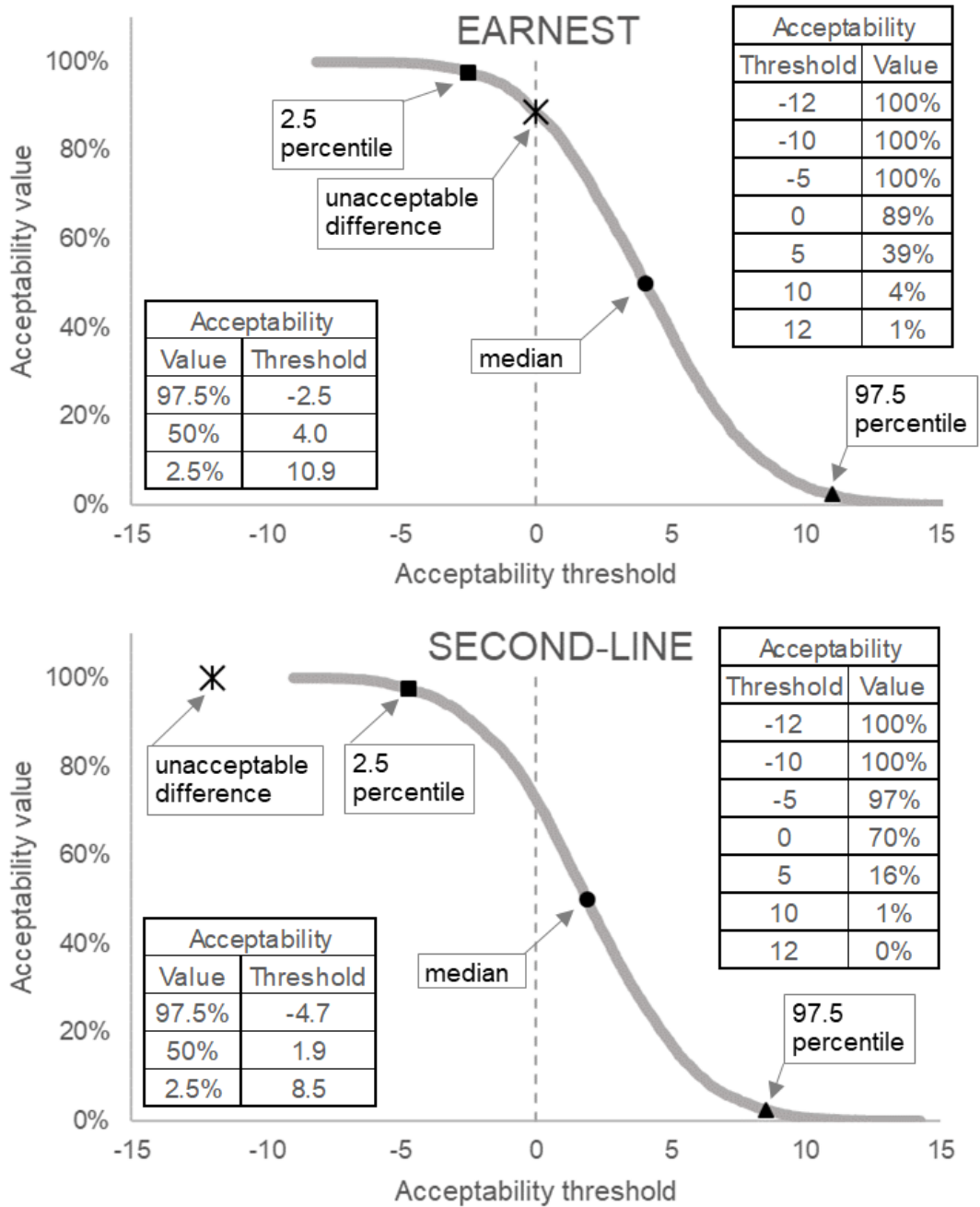

Figure 1: Acceptability Curve Estimation using Probability Above Threshold (ACCEPT) curves and tables for EARNEST and SECOND-LINE trials. An acceptability value is the probability that the true treatment difference is at least the acceptability threshold. Positive values indicate raltegravir is better than NRTI. For example, the probability that the true difference between treatments was at least 0 (i.e. that raltegravir is better than NRTI) was 89% in EARNEST and 70% in SECOND-LINE.

Median estimates, 95% credible intervals from models, and pre-specified unacceptable differences are marked.

ACCEPT can be implemented using Bayesian analysis, which provides direct estimation of the probability one treatment is better or worse than another when the prior belief of the difference between treatments is added to the analysis. The degree of prior belief, termed priors, are based on existing data and/or expert opinion. Priors are specified as a distribution over the possible values that the difference can take, with non-informative priors essentially being a flat distribution, and strongly informative priors being very concentrated around the area of highest belief. ACCEPT within a Bayesian framework is more consistent with the overall philosophy than within a frequentist framework. However, frequentist analysis using confidence curves13 14 is expected to give very similar results to Bayesian analysis with uninformative priors. In the frequentist framework, the acceptability value is the one-sided p-value for the treatment effect exceeding the acceptability threshold. An online tool enables ACCEPT for up to three trials simultaneously using summary information from frequentist or Bayesian analysis https://egon.stats.ucl.ac.uk/projects/ACCEPT/. Further details of analyses, including statistical code, is available in the supplementary information.

Using ACCEPT for trial reporting, EARNEST results would still conclude that superiority was not shown but could also include a statement such as 'ACCEPT suggested that there was a 89% probability that the true treatment difference was greater than zero (i.e. that raltegravir was better than NRTI) and 100% probability that the true treatment difference was above -5, equivalent to a 0% probability that raltegravir was worse than NRTI by at least five percentage points. There was a 39% probability raltegravir was better than NRTI by at least five percentage points.'

Similarly, reporting of SECOND-LINE results with ACCEPT would still conclude that non-inferiority was shown but could also add a statement stating 'ACCEPT suggested that there was a 70% probability of the true treatment difference being greater than zero, a 97% probability of the true treatment difference being above -5 percentage points, and a 16% probability that raltegravir was better than NRTI by at least five percentage points'.

Using ACCEPT, stakeholders requiring clear benefit of raltegravir for implementation could use acceptability thresholds of zero and above, concluding that the probability of raltegravir being better than NRTI was 89% in EARNEST and 70% in SECOND-LINE, but the probability of being more than 5 percentage points better was much lower at 39% in EARNEST and 16% in SECOND-LINE. Other stakeholders more focused on other secondary benefits of raltegravir, such as lower toxicity, could use acceptability thresholds of zero and below, concluding almost certainty of the true treatment difference being more than -5 percentage points in either trial. This allows better comparison across trials than the primary analysis alone.

Interpretation through ACCEPT has three main strengths. Firstly, it enables comparison between trials and trial types by harmonising reporting of results; the use of probabilities is straightforward, widely understood and reflects the uncertainty around the point estimate. Secondly, presentation of ACCEPT acknowledges different acceptability threshold may exist in different situations. ACCEPT allows clinicians, policymakers, and patients to make informed decisions based on their setting and individual circumstances if they feel the original choice of unacceptable difference is not appropriate for their context. Thirdly, ACCEPT moves the focus from comparison with the pre-specified unacceptable value to interpretation of the trial data. This may be especially useful for trials pre-specified as non-inferiority to reduce focus on the selected unacceptable value and in situations where restricted sample size reduces power, such as subgroup analysis and uncommon conditions; for subgroup analysis, ACCEPT can be run separately for each subgroup

Use of ACCEPT does not remove all of the concerns that can arise with non-inferiority trials, which are caused by the unacceptable difference being less than zero. Non-inferiority trials cannot always provide assurance that the new treatment has a clinically relevant effect (greater than zero) relative to placebo and so it is important to carefully assess evidence about how much better the control treatment is than placebo when selecting the pre-specified unacceptable difference/non-inferiority margin to prevent biocreep. Non-adherence in clinical trials may bias the estimate of treatment differences towards zero, especially if treatment cross-over occurs, meaning that conclusions of non-inferiority may be more likely with substantial non-adherence. Analysis of different trial populations (per protocol and intention to treat) or statistical adjustment must still be used to allow for this, but ACCEPT can help improve interpretation when comparing across different populations within a trial.

ACCEPT can be applied to historical trials or incorporated into statistical analysis plans for future analyses. ACCEPT has been previously advocated for use in clinical trials reporting, but its use has not become widespread, perhaps due to lack of common language to discuss the analyses. Increased use of a variety of different trial designs means the time is right for unified design and interpretation though ACCEPT.

## ACKNOWLEDGEMENTS

We thank Andrew Nunn, Di Gibb, Hanif Esmail, Julia Bielicki and Mike Sharland, for thoughtful comments that greatly improved the mansucript. MNC, IRW, AJC, DTD, MQ, RMT, CDT and ASW are supported by core support from the Medical Research Council UK to the MRC Clinical Trials Unit [MC_UU_12023/22 and MC_UU_00004/09]. ASW is an NIHR Senior Investigator. SC is supported by an NIHR advanced fellowship (NIHR300593).

# Supplementary information: Choice and effect of priors

We used weakly informative priors in the Bayesian analysis as truly uninformative priors on the logit scale is not possible.

The control event rate was assumed to be 75% in EARNEST and 80% in SECOND-LINE.

The control arm prior on the logit scale follows a Normal distribution with mean of the logit of the event rate and standard deviation 2.

- For EARNEST on the natural scale, this corresponds to a median of 75%, mean of 66%, interquartile range of (44%, 92%), and 95% of the distribution within (6%, 99%).
- For SECOND-LINE on the natural scale, this corresponds to a median of 80%, mean of 70%, interquartile range of (51%, 94%), and 95% of the distribution within (9%, 99.5%).

The prior for the difference between the control and treatments arms on the logit scale follow a Normal distribution with mean 0 and standard deviation 8.

- For EARNEST on the natural scale, this corresponds to median differences between arms of 0%, mean -10%, interquartile range (-50%, 19%) and 95% of the distribution within (-96%, 82%).
- For SECOND-LINE on the natural scale, this corresponds to median differences between arms of 0%, mean -14%, interquartile range (-55%, 16%) and 95% of the distribution within (-96%, 79%).

To test the influence of the priors, we compared results from frequentist glm taking margins, and the Bayesian posterior distributions from a glm. Both glms used logit link functions, binomial errors, binary outcome of success/failure and single predictor variable of arm with NRTI as the intercept. The model estimated values on the natural scale were very similar between both methods in the mean estimate, and the lower and upper 95% confidence intervals (frequentist) and Bayesian credible intervals (Bayesian).

| Trial | Estimate | Frequentist value | Bayesian value |
|---|---|---|---|
| EARNEST | Mean | 4.1% | 4.1% |
| | Lower 95% | -2.4% | -2.3% |
| | Upper 95% | 10.6% | 10.6% |
| SECOND-LINE | Mean | 1.8% | 1.7% |
| | Lower 95% | -4.7% | -5.1% |
| | Upper 95% | 8.3% | 8.4% |

# Supplementary information: R code to run analyses

```r
# load packages ------------------------------------------------
# R version 4.0.5 used
library(tidyverse) #v1.3.1
library(purrr) #v0.3.4
library(boot) # for inv.logit v1.3-27
library(margins) # for doing margins v0.3.26
library(rstanarm) # for running Bayesian v2.21.1
library(rstan) #v2.21.2
library(stringr) #v1.4.0

# my preferences ------------------------------------------------
# make sure using dplyr select
select <- dplyr::select
# graph preferences
theme_set(theme_bw(base_size = 12))
theme_update(panel.grid.major = element_blank(),
       panel.grid.minor = element_blank(),
       strip.background=element_blank())
options(tibble.print_max = 20, tibble.print_min = 50) # limit printing
# create datasets ----------------------------------------------------
# earnest
earnest <- tibble(
  arm = c(rep("NRTI", 426), rep("Rtvr", 433)),
  result = c(rep(1, 255), rep(0, 426-255), rep(1, 277), rep(0, 433-277))
)
# checking success rate
earnest %>%
  group_by(arm) %>%
  summarise(n = n(),
       n_successes = sum(result),
       mean = mean(result))
```

```r
# second line
secondline <- tibble(
  arm = c(rep("NRTI", 271), rep("Rtvr", 270)),
  result = c(rep(1, 219), rep(0, 271-219), rep(1, 223), rep(0, 270-223))
)
secondline %>%
  group_by(arm) %>%
  summarise(n = n(),
        n_successes = sum(result),
        mean = mean(result))
# frequentist analysis ---------------------------------------------
# model and margins in frequentist
mfreq_earnest <- glm(result ~ arm, data = earnest, family = binomial())
summary(mfreq_earnest)
summary(margins(mfreq_earnest))

mfreq_secondline <- glm(result ~ arm, data = secondline, family = binomial())
summary(mfreq_secondline)
summary(margins(mfreq_secondline))
# priors -------------------------------------------
# 1.96
critval = qnorm(0.975,mean=0,sd=1)
# useful function
logodds <- function(x) log(x/(1-x))
### earnest
# assumed rates in control arm and delta/NI margin
p1_earnest <- 0.75

# p1 intercept prior
intercept_earnest_mean <- logodds(p1_earnest)
intercept_earnest_sd <- 2
# weakly informative beta prior
noeffect_earnest_mean <- 0
noeffect_earnest_sd <- 8
### secondline
```

```r
# assumed rates in control arm and delta/NI margin
p1_secondline <- 0.8
# p1 intercept prior
intercept_secondline_mean <- logodds(p1_secondline)
intercept_secondline_sd <- 2
# weakly informative beta prior
noeffect_secondline_mean <- 0
noeffect_secondline_sd <- 8
# stan: earnest weakly informative beta prior --------------------------------------------------

mbayes_earnest <- stan_glm(result ~ arm, family = binomial, data = earnest,
                 prior = normal(location = noeffect_earnest_mean, scale = noeffect_earnest_sd),
                 prior_intercept = normal(location = intercept_earnest_mean, scale = intercept_earnest_sd)
)
# outputs
prior_summary(mbayes_earnest)
posterior_vs_prior(mbayes_earnest)
stan_trace(mbayes_earnest)
# summary
summary(mbayes_earnest)
# posterior check (dark line is the data)
pp_check(mbayes_earnest)
# shiny stan
# launch_shinystan(mbayes_earnest)
stan_hist(mbayes_earnest)

# get posteriors iterations and put onto natural scale
post_earnest <- as.data.frame(mbayes_earnest) %>%
  rename(
    intercept_logit = `(Intercept)`,
    arm_effect_logit = armRtvr) %>%
  mutate(
    nrti_natural = invlogit(intercept_logit),
    rtvr_natural = invlogit(intercept_logit + arm_effect_logit),
```

```r
    diff_natural = rtvr_natural - nrti_natural
  ) %>%
  mutate(type= "earnest")
# summaries on natural scale
summary(post_earnest$nrti_natural)
summary(post_earnest$rtvr_natural)
summary(post_earnest$diff_natural)
quantile(post_earnest$diff_natural, c(0.025, 0.975))

# stan: secondline weakly informative beta prior -------------------------------------------------
mbayes_secondline <- stan_glm(result ~ arm, family = binomial, data = secondline,
                   prior = normal(location = noeffect_secondline_mean, scale = noeffect_secondline_sd),
                   prior_intercept = normal(location = intercept_secondline_mean, scale = intercept_secondline_sd)
)
# outputs
prior_summary(mbayes_secondline)
posterior_vs_prior(mbayes_secondline)
stan_trace(mbayes_secondline)
# summary
summary(mbayes_secondline)
# posterior check (dark line is the data)
pp_check(mbayes_secondline)
# shiny stan
# launch_shinystan(mbayes_secondline)
stan_hist(mbayes_secondline)
# get posteriors iterations and put onto natural scale
post_secondline <- as.data.frame(mbayes_secondline) %>%
  rename(
    intercept_logit = `(Intercept)`,
    arm_effect_logit = armRtvr) %>%
  mutate(
    nrti_natural = invlogit(intercept_logit),
    rtvr_natural = invlogit(intercept_logit + arm_effect_logit),
```

```r
    diff_natural = rtvr_natural - nrti_natural
  ) %>%
  mutate(type= "secondline")

# summaries on natural scale
summary(post_secondline$nrti_natural)
summary(post_secondline$rtvr_natural)
summary(post_secondline$diff_natural)
quantile(post_secondline$diff_natural, c(0.025, 0.975))
# posteriors ------------------------------------------------

posteriors <- post_earnest %>%
  bind_rows(post_secondline)

# make trial names nice
posteriors <- posteriors %>%
  mutate(trial = ifelse(str_detect(type, "earnest"), "EARNEST",
                ifelse(str_detect(type, "secondline"), "SECOND-LINE", NA)))

# make a dataset with expected and unacceptable values and dtd values
values <- tibble(
  trial = c("EARNEST", "EARNEST", "SECOND-LINE", "SECOND-LINE"),
  name_long = c("Unacceptable difference", "Expected difference", "Unacceptable difference", "Expected difference"),
  value_ppts = c(0, 10, -12, 0))

# Looking at posteriors ------------------------------------------------
# posteriors with ecdf
posteriors <- posteriors %>%
  group_by(trial) %>%
  # multiple diff_natural by 100 to get percentage points
  mutate(diff_natural_ppts = diff_natural * 100) %>%
  arrange(trial, diff_natural_ppts) %>%
  # get cumulative distribution - need probabilty at least threshold value so do one minus
```

```r
  mutate(effect_ecdf = 1- ecdf(diff_natural_ppts)(diff_natural_ppts))

# get summary stats - need to do one minus to match above
summ <- posteriors %>%
  group_by(trial) %>%
  summarise(
    mean = mean(diff_natural_ppts),
    q50 = quantile(diff_natural_ppts, 0.5),
    q025 = quantile(diff_natural_ppts, 0.025),
    q975 = quantile(diff_natural_ppts,c(0.975))
  )
# make quantiles long to plot
summ_long <- summ %>%
  select(-mean) %>%
  pivot_longer(q50:q975) %>%
  mutate(quantile = ifelse(name == "q50", 0.5,
                    ifelse(name == "q025", 0.025,
                    ifelse(name == "q975", 0.975, NA))),
         name_long = ifelse(name == "q50", "50 percentile (median)",
                     ifelse(name == "q025", "2.5 percentile ",
                     ifelse(name == "q975", "97.5 percentile", NA)))) %>%

  # need to do one minus to match above
  mutate(quantile = 1- quantile)
## work on the unacceptable and expected difference values - I want small lines added to the graphs at the correct point on the y axis.
# join posteriors and values together - this will double the size of the dataset so each posterior in each trial is repeated for each didfference
post_values <- left_join(
  posteriors %>%
    select(trial, diff_natural_ppts, effect_ecdf),
  values, by = "trial"
)
# group by the trial and name_long and then find the closest effect_ecdf to the expected and unacceptable differences
post_values <- post_values %>%
```

```r
  group_by(trial, name_long) %>%
  mutate(diff_abs = abs(diff_natural_ppts - value_ppts)) %>%
  mutate(minflag = ifelse(diff_abs == min(diff_abs), 1, 0)) %>%
  filter(minflag == 1) %>%
  select(trial, name_long, value_ppts, effect_ecdf)

# put summ_long and post_values together
valuestoplot <- summ_long %>%
  select(trial, name_long, xvalue = value, yvalue = quantile) %>%
  bind_rows(
    post_values %>%
      select(trial, name_long, xvalue = value_ppts, yvalue = effect_ecdf)
  ) %>%
  mutate(
    name_long = factor(name_long, levels = c("2.5 percentile ", "50 percentile (median)", "97.5 percentile", "Unacceptable difference", "Expected difference"))
  )
# plot ecdf
p1<- ggplot(posteriors, aes(x = diff_natural_ppts, y = effect_ecdf)) +
  geom_vline(xintercept = 0, linetype =3)+
  geom_line(size = 2, colour = "gray78") +
  facet_wrap(~trial) +
  geom_point(data = valuestoplot,
             aes(x = xvalue, y = yvalue, shape = name_long, size = name_long)) +
  # scale_shape_manual(values = c(15, 16, 17, 124, 124)) +
  scale_shape_manual(values = c(15, 16, 17, 91, 124)) +
  scale_size_manual(values = c(2.5, 2.5, 2.5, 5, 5)) +
  labs(x = 'Acceptability threshold', y = "Acceptability value") +
  scale_y_continuous(labels = scales::percent_format(accuracy = 1)) +
  theme(legend.position = "bottom", legend.title = element_blank())
p1
# Probability tables -------------------------------------------
post_summary <- posteriors %>%
  group_by(trial) %>%
  summarise(
```

```
    accthreshold_minus12 = mean(diff_natural_ppts > -12),
    accthreshold_minus10 = mean(diff_natural_ppts > -10),
    accthreshold_minus5 = mean(diff_natural_ppts > -5),
    accthreshold_0 = mean(diff_natural_ppts > 0),
    accthreshold_5 = mean(diff_natural_ppts > 5),
    accthreshold_10 = mean(diff_natural_ppts > 10),
    accthreshold_12 = mean(diff_natural_ppts > 12)) %>%
  ungroup() %>%
  pivot_longer(starts_with("accthreshold_")) %>%
  # extract numebrs from acc thes and make negative if necesary
  mutate(acceptability_threshold = as.numeric(str_extract(name, "(\\d)+")),
         acceptability_threshold = ifelse(str_detect(name, "minus"), -1*acceptability_threshold, acceptability_threshold))

post_summary_table <- post_summary %>%
  mutate(value = paste0(signif(value * 100, 2), "%")) %>%
  pivot_wider(id_cols = acceptability_threshold, names_from = trial)

post_summary_table
```